\documentstyle[preprint,prb,aps]{revtex}
\begin{document}
\draft
\title{Titanium additions to $MgB_2$ conductors}
\author{N. E. Anderson, Jr., W. E. Straszheim, S. L. Bud'ko, P. C. Canfield, and D. K. Finnemore}
\address{Ames Laboratory, U.S. Department of Energy and Department of Physics and Astronomy\\
Iowa State University, Ames, Iowa 50011}
\author{Raymond J. Suplinskas}
\address{Specialty Materials, Inc.\\
1449 Middlesex Street, Lowell MA 01851}
\date{\today}
\maketitle
\begin{abstract}
A series of doping experiments are reported for $MgB_2$ conductors that have been synthesized 
using doped boron fibers prepared by chemical vapor deposition (CVD) methods.  Undoped $MgB_2$ samples prepared from 
CVD prepared fibers  
consistently give critical current densities, $J_c$, in the range of $500,000~
A/cm^2$ in low field at $5~K$.  These values fall by a factor of about $100$ as the magnetic 
field increases to $3~T$.  For heavily Ti-doped boron fibers where the B/Ti ratio is comparable 
to 1, there is a substantial suppression of both $T_c$, superconducting volume fraction, and $J_c$ values. 
If, however, a sample with a few percent Ti in B is deposited on a carbon substrate and reacted at $1100^{\circ }C$, then $T_c$ is suppressed only a couple of degrees and 
critical current densities are found to be  $J_c\sim 2-5\times 10^6~A/cm^2$ for superconducting 
layers ranging from $4-10~\mu m$ thick.  These materials show $J_c$ values over $10,000 A/cm^2$ at $25~K$ and $1.3~T$.

\end{abstract}


\section{Introduction}

Shortly after the discovery of superconductivity in $MgB_2$\cite {1}, small wire segments 
of this compound were prepared by reacting commercial $B$ fibers 
in a $Mg$ vapor.\cite {2} Typically these wire segments of pure $MgB_2$ will carry 
current densities, $J_c$, in the range of a few hundred thousand $A/cm^2$ in self field and 
$5~K$. Fully reacted wire segments were consistently found to show critical current densities, 
$J_c$ ranging from $100~kA/cm^2$ to $500~kA/cm^2$ at $5~K$ and 
magnetic fields in the range of 
$0.1~T$ with a drop by a factor of about $100$ as the magnetic field increases to $3~T$.  

In light of the early success of  the $Mg$ vapor diffusion method of sample preparation, a more detailed 
study of the  of the $MgB_2$ growth process\cite {3}was undertaken and has shown  
that as $Mg$ diffuses into $B$, a wall of $MgB_6$ sweeps across the boron 
fiber, followed by a wall of $MgB_4$, and finally a wall of $MgB_2$.   Energy dispersive spectra ($EDS$) in a scanning 
electron microscope ($SEM$) have shown that the $Mg$ content is rather uniform across any given 
phase and rises abruptly at the phase boundary.  For example, in a region containing both 
 $MgB_4$ and $MgB_2$,  the $Mg$ concentration as measured by a $1~\mu m$ diameter 
EDS beam is rather uniform in the   
$MgB_4$ region and then rises abruptly at the interface between the $MgB_4$ and $MgB_2$ 
regions within the resolution of the $EDS$ probe of $1$ to $2~\mu m$. Then, it is again rather uniform through the $MgB_2$ region.  

Very successful $Ti$-doping experiments were reported by Zhao et al.\cite {4} using a 
solid state reaction method in which powders of $Mg$, $Ti$, and $B$ were mixed, pressed, 
and reacted on an $MgO$ plate in flowing $Ar$ gas.  Best results were found for a $Ti/Mg$ ratio of about $0.1$ which showed $J_c\sim 2\times 10^6~A/cm^2$ at $5~K$ and zero field.  Another recent paper by Wang et al.\cite {5} shows that $Y_2O_3$ nano-particles can 
greatly enhance the irreversibility field  and $J_c$ performance of $MgB_2$ in the $3-4~T$ range for samples prepared by a powder-in-tube method.  In an alternate approach, Komori et al.\cite {6} have used a pulsed laser deposition method to prepare $0.5 ~\mu m$ thick films of $MgB_2$ 
on a buffered flexible substrate that carry about $100,000~A/cm^2$ at $4.2~K$ and $10~T$.  
These thick films are special because the grain size is in the nanometer range and $J_c$ 
remains high out to $10~T$.

The purpose of the work reported here is to explore the use of continuous $CVD$ methods 
to dope the $MgB_2$ phase.  It is important to 
increase both the upper critical field, $H_{c2}$ 
and the critical current densities in high field by shortening the mean-free-path of the electrons 
and by adding suitable pinning centers for the superconducting vortices.  Impurities tend to 
suppress the transition temperature, $T_c$, so it is important to find a trade-off such that the 
$T_c$ is not reduced too much, but the critical field and critical current are substantially 
 improved.  The goal is to find suitable impurities and convenient methods to uniformly disperse them so that the $J_c$ values increase with relatively little suppression of $T_c$.  Emphasis is on 
chemical vapor deposition processing of the $B$ and vapor diffusion of $Mg$ to form the $MgB_2$ conductor.

\section{Experiment}

A variety of methods have been used to introduce both point defects and small precipitates that 
will both shorten the electronic mean free path and provide small precipitates.  
To produce $Ti$ impurites, the  $Ti$ and $B$ are 
co-deposited in a reel-to-reel CVD chamber.\cite {7}  Two different substrates were used 
for the co-deposition of  $B$ and $Ti$.  In some experiments, the substrate is a $W$ wire.  
A $15~\mu m$ diameter $W$ continuous filament enters the $CVD$ chamber 
in which a mixture of $BCl_3$ and $H_2$ are flowing.  The conductor 
is electrically heated to peak temperatures of $1100^{\circ }C$ and either pure $B$ or 
$B$-$Ti$ mixtures can be deposited.  
Titanium is added by bubbling the hydrogen component through $TiCl_4$ held at a 
temperature near $0^{\circ }C$.  The resulting partial pressure of $TiCl_4$ is approximately 
$2\times 10^{-3}$ bar.  The $Ti$ level is set by the temperature of the $TiCl_4$.   In other experiments, the substrate for boron deposition 
was a $75~\mu m$ diameter $SiC$ filament coated with several micrometers of glassy carbon 
known commercially as $SCS-9a$.  The $CVD$ coated fibers were spooled continuously 
on emerging from the reactor.  After co-deposition of $B$ and $Ti$, X-ray patterns show both  $TiB$ and $TiB_2$ in the fibers.  The boron is transformed to $MgB_2$ 
by immersion in $Mg$ vapor as previously reported.\cite {2}.  For the $Ti$-doped samples, 
more details of the structure and $Ti$ distribution has been presented elsewhere.\cite {8}

Critical current values were determined from magnetization hysteresis loops using the Bean Model for cylindrical geometry\cite {4} by 

$$J_c~={ ~{1.5 \Delta M }\over {r_o[1~-~r_i^3/r_o^3]}}~~\eqno {1}$$

\noindent where $\Delta M$ in A/m is the difference in magnetization for field increasing and field decreasing, 
$r_o$ is the outer radius of  the superconductor, and $r_i$ is the inner radius measured in m.  Magnetization measurements were made in a Quantum Designs SQUID magnetometer.  Both 
$M~vs~T$ and the $M~vs~H$ hysteresis loops were measured sequentially on the same sample.

\section{Results and discussion}

To compare relative performance of various samples, a reference sample of 
 pure $MgB_2$ sample was made by reacting $100~\mu m$ diameter $CVD$ boron 
fibers in a $Ta$ container at $950^{\circ }$C for $2~h$.\cite {2} The resulting magnetization 
curve in an applied field of $50~Oe$, is illustrated in the inset of Fig. 1 and shows the onset of 
superconductivity at $39.2~K$ and a transition width of a few tenths of a Kelvin.  
In a separate control measurement, a sample with 
a pure boron layer $25~\mu m$ thick was deposited on a carbon coated $SiC$ fiber and 
reacted in $Mg$ vapor as above.\cite {2}  As reported elsewhere,\cite {8} the $J_c~vs.~H$ 
curves of this sample are identical to the sample with pure $B$ deposited on $W_2B_5$ to a 
thickness of $100~\mu m$.  Changing the substrate from $W_2B_5$ to carbon coated $SiC$ 
and changing the thickness from the $25~\mu m$ to $100~\mu m$ does not change 
the $J_c~vs.~H$ behavior of pure $MgB_2$.  These results would appear to imply 
that $C$ is not diffusing from the $C$ substrate into the $MgB_2$ during reaction at 
$950^{\circ }$C for $2~h$ to a level that will 
have a significant effect on $J_c$.

\subsection {Heavy $Ti$-doping}

Doping with $Ti$ has been done at both the $50\%$ and $5\%$ levels.  To get a broad perspective, a sample with a high $Ti$ level was prepared.and converted to $MgB_2$ by reacting at $950^{\circ }$C for $2~h$.  This high $Ti$ level sample was deposited on a $15~\mu m$ 
$W$ substrate similar to the standard pure $MgB_2$, but the diameter was $80~\mu m$ 
rather than $100~\mu m$ as in the pure sample. 
The resulting sample had a $Mg/Ti$ ratio of about 1 as determined by the $EDS$.  , 
As shown by the upper curve in the inset of Fig. 1, the low field magnetization has a small diamagnetic drop at $\sim 38~K$ and a larger drop in the region of $25~K$. An energy dispersive X-ray (EDS) study in the scanning 
electron microscope (SEM) showed comparable amounts of $Ti$ and $Mg$ which is consistent 
with the Meissner screening fraction of about $25~G$ in an external field of $50~Oe$.  As 
shown by the solid squares of  Fig. 1, the pure $MgB_2$ sample at $5~K$ shows 
$J_c\sim 650,000~A/cm^2$ at zero field and $J_c\sim 10,000~A/cm^2$ at $1.5~T$.  At $25~K$, shown by the solid triangles, $J_c\sim 300,000~A/cm^2$ at zero field and $J_c\sim 10,000~A/cm^2$ at $0.7~T$.  
For the $B/Ti$ ratio approximately equal to 1 sample, 
$J_c$ values at $5~K$ are about $200,000~A/cm^2$ 
at zero applied field, but drop below $10,000~A/cm^2$ in just a few tenths of a Tesla.  This sample clearly has far too much $Ti$.

\subsection {Light $Ti$-doping}

Samples 
with nominally a $Ti/B$ ratio of $0.05$ were all deposited onto $SiC$ substrates coated 
with glassy carbon
so these samples have both $Ti$ and possibly some small amount of 
$C$ impurities. A series of  $EDS$ measurements in the  $SEM$ 
show that different samples with boron thicknesses of $4$, $6$, and $10~\mu m$  
 have somewhat different $Ti$ concentrations. 
These $EDS$ studies with a sampling 
beam diameter of $\sim 1~\mu m$ indicate that the  $4~\mu m$, the $6\mu m$, and 
 the $10~\mu m$ thick samples have $Ti$ contents roughly in the ratio of $2:1.5:1$.  The 
$10~\mu m$ thick sample has roughly $9$\% as many $Ti$ atoms as $Mg$ atoms.  The $Ti$ 
and $Mg$ peaks are strong and isolated so the result is relatively easy to measure.  A 
systematic study of the causes for the variation in the $Ti$ content with $B$ thickness has 
not yet been undertaken, but it is most likely related to the fact that the temperature dependence 
of  $Ti$ is deposition differs from the temperature dependence of $B$ deposition.      

Low field  magnetization characteristics at $50~Oe$ are  shown in the inset of Fig. 2  for the $4\mu m$ thick sample that has been reacted in $Mg$ vapor 
 at $950^{\circ }$C for $1~h$ and $2~h$.  The onset of Meissner screening is systematically suppressed to values 
near $36~K$ and crosses $50~Oe$ at about $33~K$.  For these samples with a few 
micrometers of superconductor on a $75~\mu m$ diameter  $SiC$ substrate, there is some difficulty determining both the mass of superconductor present.  In addition, the reacted 
samples are not completely straight or really well aligned with the magnetic field so the  demagnetizing factor for the sample has some uncertainty.  Thus, the absolute magnitude of the 
$4\pi M$ values could be in error by $30$ to $40$ percent.   All samples were treated in 
the same manner, however, so sample to sample comparisons are more reliable than the 
absolute value of the magnetization.  The pure 
$MgB_2$ sample of Fig. 2 and Fig. 3 is a $100~\mu m$ diameter boron fiber on a 
$15~\mu m$ diameter $W_2B_5$ substrate that has been reacted in $Mg$ vapor at 
$950^{\circ }C$ for $2~h$.

The semi-logarithmic plot of $J_c$ vs field for these two samples at $5~K$ are a factor of $8$ higher than pure 
$MgB_2$ at low field.  Both the $1~h$ reaction and the $2~h$ reaction samples have essentially the same 
$J_c$ dropping from 
about $5,000,000~A/cm^2$ at $0.1~T$ to about $50,000~A/cm^2$ at $3~T$ as shown in 
Fig. 2.   Above $2~T$, the improvement is very substantial.

Because short reaction times are usually desirable 
in any commercial process, a series of samples ranging in thickness 
from $4-10~\mu m$ were deposited on a $C$ substrate and reacted at $1100^{\circ }$C for 
$15~min$.  The transition temperature again is suppressed to about $35~K$ as shown in 
the inset of Fig. 3.  
With these reaction conditions, the value of $J_c$ in zero applied field systematically drops from 
about $4,000,000~A/cm^2$ for a $4~\mu m$ thick layer to about $2,000,000~A/cm^2$ for a 
$10~\mu m$ thick layer as shown in Fig. 3.  If these data are plotted on a semi-logarithmic scale to 
emphasize the lower $J_c$ values at high magnetic field, this trend with thickness is reversed. 
In the region of  $3~T$, the $10~\mu m$ thick sample shown by the solid triangles in Fig. 4 
carries more than $100,000~A/cm^2$ and 
the $4~\mu m$ thick sample shown by the solid squares carries about $10,000~A/cm^2$.  If 
the temperature is raised to $25~K$, the $10~\mu m$ thick sample has $J_c\sim 
600,000~A/cm^2$ in zero applied field and about $J_c\sim 10,000~A/cm^2$ at $1.3~T$ as 
shown on Fig. 4.  .

\subsection {Optimization of $Ti$-doped samples}

Optimization of the $J_c$ performance as a function of reaction times was performed 
at two different temperatures of $1100^{\circ }C$ and $950^{\circ }C$.  
Here, the $25~K$ data for $J_c$ are shown because this is the most likely 
operating temperature.  As shown in Fig. 5a, the optimum reaction time at $1100^{\circ }C$ is in 
the range of $15$ to $30$ minutes, and the samples show $J_c~\sim 10,000~A/cm^2$ at 
$1.2~T$.  As shown in Fig. 5b, the optimum reaction time at $950^{\circ }C$ is about $2~h$, 
and the samples show $J_c~\sim 10,000~A/cm^2$ at $1.3~T$.  Our assumption has been that 
the degradation of $J_c$ and $T_c$ at long times may arise from the diffusion of carbon from the substrate 
into the $MgB_2$.  It is difficult, however, to determine the carbon content in these layers of 
$MgB_2$ that are $4$ to $10~\mu m$ thick and are on a $C$ substrate.  The $EDS$ carbon peak is a shoulder on the $B$ peak 
and is small compared to the $Ti$ peak for all temperature and times studied for the $10~\mu m$ 
thick sample.  If you were very optimistic about the data interpretation, the carbon shoulder may 
grow to a value of $10$ percent of the titanium peak for the sample reacted 
at $1100^{\circ }C$ and $45~min$ where there is significant degradation of $J_c$. 
 This degradation of $J_c$, however,  may also come from a rearrangement of the $Ti$ as well. 
Further work needed to improve the determination of the
 carbon content and the causes for the observed changes in $J_c$ of the superconductor.

\section{Conclusions}

Titanium impurities co-deposited with the boron in a $CVD$ process is a very effective 
method to raise  
the critical current performance in the range $3~T$ at $5~K$ and in the range of $1.3~T$ at 
$25~K$.  The value of  $T_c$ is suppressed only $2~K$ or $3~K$ and there is a very substantial 
enhancement of $J_c$. These $CVD$ methods are an effective way to distribute the $Ti$ impurities and can be 
used to produce conductors with  $J_c$ values in the range of $10,000~A/cm^2$ 
at $25~K$ and $1.3~T$.

\section{Acknowledgments}
Work is supported by the U.S. Department of Energy, Basic Energy Sciences, 
Office of Science, through the Ames Laboratory under Contract No. W-7405-Eng-82.

\vfil\eject

\begin{figure}
\caption{Values of $J_c$ for  pure $MgB_2$ and a sample with a $Ti/Mg$ ratio of about $1$. 
The inset shows magnetization curves in $50~Oe$. } 
\end{figure}

\begin{figure}
\caption{Semi-logarithmic plot of values of $J_c$ for $Ti$-doped $MgB_2$ reacted at 
$950^{\circ }$C.  The inset 
shows magnetization vs. temperature data in the $30-40~K$ range for these samples in $50~Oe$.  At low $T$, the data level out at about $100-150~G$.} 
\end{figure}

\begin{figure}
\caption{Thickness variation of $J_c$ for samples doped with both $Ti$ and $C$ and reacted 
at $1100^{\circ }$C.  Inset shows 
magnetization for these samples in $50~Oe$.  At low $T$, the data level out at about 
$75-100~G$.} 
\end{figure}

\begin{figure}
\caption{Semi-logarithmic plot to compare the high field and low field variation in $J_c$ with thickness.  For the $10~\mu m$ sample, the $5~K$ data are shown by the solid triangles with 
the point up, 
and the $25~K$ data are shown by the solid triangles with the point down.} 
\end{figure}

\begin{figure}
\caption{Performance characterics for $10~\mu m$ thick samples reacted at (a) 
$1100^{\circ }C$ and (b) $950^{\circ }C$.} 

\end{figure}
\vfil\eject

\end{document}